\documentclass{svproc}
\usepackage{url}

\usepackage{amsmath}
\usepackage{amsfonts}
\usepackage{amssymb}
\usepackage[english]{babel}
\usepackage{graphicx}
\usepackage[utf8]{inputenc}

\begin{document}
\mainmatter              
\title{Adjoint-based Identification of Sound Sources for Sound Reinforcement and Source Localization}
\titlerunning{Adjoint Sound}  
%
\author{Mathias Lemke \and Lewin Stein}
\authorrunning{Mathias Lemke et al.} 
%
\tocauthor{Mathias Lemke and Lewin Stein}
\institute{Institut für Strömungsmechanik und Technische Akustik, \newline Technische Universität Berlin, Germany\\
\email{mathias.lemke@tnt.tu-berlin.de}}

\maketitle              

\begin{abstract}
The identification of sound sources is a common problem in acoustics.
Different parameters are sought, among these are signal and position of the sources.
We present an adjoint-based approach for sound source identification, which employs computational aeroacoustic techniques.
Two different applications are presented as a proof-of-concept: optimization of a sound reinforcement setup and the localization of (moving) sound sources.

\end{abstract}

\section{Introduction}

\label{sec:Introduction}
A common issue in acoustics is the identification of fixed or moving sound sources.
In general, several parameters have to be determined; among these are the source signal and the position of the sources.
This general problem occurs in many applications, from environmental to industrial acoustics.

In this contribution, we discuss an adjoint-based approach for sound source identification.
The time-domain method is based on the (adjoint) Euler equations, which are solved by means of computational aeroacoustic techniques (CAA). 
The approach allows considering complex base flows, such as non-homogeneous base flow, thermal stratification as well as complex geometries.

Adjoint-based methods have been used in the field of fluid mechanics for decades.
They have proven to be an effective approach for the analysis of flow configurations and determining optimal model parameters in various applications \cite{GilesPierce2000}.
Adjoint-based techniques are used to optimize flow configurations by means of geometry modifications \cite{Jameson1995} or for active flow control applications \cite{CarnariusThieleOzkayaNemiliGauger2013}.
They are applied for the analysis and optimization of reactive flow configurations \cite{LemkeReissSesterhenn2014,LemkeCaiReissPitschSesterhenn2018} and data assimilation applications \cite{YangRobinsonHeitzMemin2015,LemkeSesterhenn2016,GrayLemkeReissPaschereitSesterhennMoeck2017}.
Furthermore, they are employed in the field of aeroacoustics \cite{Freund2011,SchulzeSchmidSesterhenn2011} and sound reinforcement applications \cite{LemkeStraubeSchultzSesterhennWeinzierl2017,SteinStraubeSesterhennWeinzierlLemke2019}.

Here, we restrict ourselves to two applications from the areas of sound reinforcement and sound source localization with generic setups as a proof-of-concept.

In the context of sound reinforcement, line arrays are used for the synthesis of sound fields.
The identification of the geometric arrangement and the electronic drive of the loudspeaker cabinets to optimally (re-)produce a sound field is an ill-posed, inverse problem.
Typically frequency domain approaches are employed \cite{Feistel2013,ThompsonLuzarraga2013}.

For the localization of moving and non-moving sound sources, usually, microphone array methods like beam-forming are used. 
Depending on the specific task, different algorithms, working in the time domain or in the frequency domain, are applied. 
See \cite{Merino2019} for a recent overview. 

The manuscript is organized as follows:
In Sec.~\ref{sec_adjoint}, the adjoint approach is introduced, and the adjoint \textsc{Euler} equations are derived.
After a short description of the numerical implementation in Sec.~\ref{sec_caa_framework}, the derived framework is employed for an application in the context of sound reinforcement in Sec.~\ref{sec_sound_reinforcement}.
The applicability of the approach for localization of sound sources is discussed in Sec.~\ref{sec_source_localization}.

\section{Adjoint Approach \label{sec_adjoint}}
\subsection{General Adjoint Equations}
Adjoint equations can be derived in different ways, e.g.,~the continuous or the discrete approach.
Despite different discretizations, the approaches are consistent and applicable, see \cite{GilesPierce2000} for a discussion.
In addition, automatic differentiation techniques are used to create adjoint codes from existing simulation programs.
Recently, a mode-based approach to derive adjoint operators was presented \cite{ReissLemkeSesterhenn2018} as an enhancement of a direct operator construction method \cite{LemkeCaiReissPitschSesterhenn2018}.

Here, the adjoint equations are introduced in a discrete manner. 
A matrix-vector notation is used, in which the vector space is the full solution in space and time.
The section is based on \cite{GilesPierce2000,Lemke2015}.

In general, the adjoint equations arise by a scalar-valued objective function $J$, which is defined by the user and encodes the target of the analysis, e.g.,~an optimization.
It is given by the scalar product between a weight vector $g$ and a system state vector $q$
\begin{equation}
  J = g^\text{T}q.
\end{equation}
The system state $q$ is the solution of the governing system
\begin{equation}
  Aq = s \label{eq_linear_system}
\end{equation}
with $A$ as governing operator and $s$ as right-hand side forcing. 
In order to optimize $J$ by means of $s$ in terms of a brute-force approach, the governing equation has to be solved for all possible $s$.
 
Instead, to reduce the computational effort, the adjoint equation can be used
\begin{equation}
  A^\text{T}q^* = g ,\label{eq_adjoint_system_1}
\end{equation}
with the adjoint variable $q^*$.

With
\begin{equation}
  J = g^\text{T}q = \left( A^\text{T}q^* \right)^\text{T}q={q^*}^\text{T}Aq = {q^*}^\text{T}s
\end{equation}
a formulation is found, which enables the computation of the objective $J$ without solving the governing system for every possible $s$. 
With the solution of the adjoint equation, the objective can be calculated by a scalar product.
Thus, the adjoint approach enables efficient computation of gradients for $J$ with respect to $s$.

\subsection{Adjoint Euler equations for Acoustic Applications}
The section is based on \cite{Lemke2015,SteinStraubeSesterhennWeinzierlLemke2019}.
The objective function $J$ is defined in space and time with $\mathrm d \Omega = \mathrm d x_i \mathrm d t$ in the whole computational domain:
\begin{equation}
  J = \frac{1}{2} \iint \left( q - q^{\mathrm{target}} \right)^2 \mathrm d \Omega.
\end{equation}
The variable $q$ contains the full state $q = [\varrho, u_j, p]$ of the system governed by the \textsc{Euler} equations. 
Therein, $\varrho$ denotes the density, $u_j$ the velocity in the direction $x_j$, and $p$ the pressure.

For the following aeroacoustic analyses the evaluation of the objective function is restricted to the pressure, resulting in
\begin{equation}
  J = \frac{1}{2} \iint \left( p - p^{\mathrm{target}} \right)^2 \sigma ~\mathrm d \Omega. \label{eqn_objective}
\end{equation}
The additional weight $\sigma(x_i,t)$ defines where and when the objective is evaluated.
In general, the objective function has to be supplemented by a regularization term, which is omitted here for the sake of clarity.
The target $p^{\mathrm{target}}$ is application-specific.
For optimization tasks, as presented in Sec.~\ref{sec_sound_reinforcement}, it is defined corresponding to a desired sound field, e.g.,~optimal listening experience for the auditorium of an open-air concert.
For the source localization application presented in Sec.~\ref{sec_source_localization}, the target pressure is defined by microphone measurements.
The microphone positions are included by means of the weight function $\sigma$.
In both cases, a minimum of $J$ is desired.

This minimum is to be achieved under the constraint that the \textsc{Euler} equations
\begin{eqnarray}
\label{eq_euler_equations_vectorwise}
	\partial_t     	\begin{pmatrix} \varrho 	\\ \varrho u_j 				\\ \frac{p}{\gamma-1} 		\end{pmatrix} 
 + 	\partial_{x_i} 	\begin{pmatrix} \varrho u_i 	\\ \varrho u_i u_j + p \delta_{ij} 	\\ \frac{u_i p\gamma}{\gamma-1} \end{pmatrix}  \nonumber
 - u_i 	\partial_{x_i} 	\begin{pmatrix} 0	\\ 0 					\\ p 				\end{pmatrix} =  
			\begin{pmatrix} 0 		\\ 0 					\\ s_p 				\end{pmatrix},
\end{eqnarray}
with $\gamma$ as heat capacity ratio, are fulfilled.
The summation convention applies.
For details on the formulation, in particular, the reformulation of the energy equation in terms of pressure, see \cite{LemkeReissSesterhenn2014}.

To ease the derivation, the above system of partial differential equations is abbreviated by
\begin{equation}
  E(q) = s. \label{eq_direct_system}
\end{equation}

The terms $s = [0,0,s_p]$ on the right side of the \textsc{Euler} equations characterize monopole sound sources, which allow controlling the system state, respectively, the solution of the equations.
In general, also mass and momentum source terms could be considered.
The overall goal is to obtain a solution of the \textsc{Euler} equations, which reduces the objective \eqref{eqn_objective} by adapting $s$.
An optimization of $s$ corresponds to an optimization of the loudspeakers' output signals.

To use the adjoint approach for optimizing $s$, the objective function \eqref{eqn_objective} and the governing system \eqref{eq_euler_equations_vectorwise} have to be linearized.
This results in
\begin{equation}
  \delta J = \iint \underbrace{\left( q - q^{\mathrm{target}} \right) \sigma}_{=g}  \delta p \mathrm d \Omega, 
\end{equation}
and
\begin{equation}
  E_{\mathrm{lin}} \delta q = \delta s.
\end{equation}
The weight $g = (q - q^{\mathrm{target}})\sigma$ encodes the difference between the current numerical solution and the target field.
Here, it is evaluated only in terms of pressure, as discussed above.
Combining the linearized system and the objective in a \textsc{Lagrangian} manner leads to
\begin{eqnarray}
  \delta J &=& g^\text{T} \delta q - {q^*}^\text{T} \underbrace{\left( E_{\mathrm{lin}} \delta q - \delta s \right)}_{=0} \\\nonumber
           &=& {q^*}^\text{T}\delta s + \delta q^\text{T} \left(g - E_{\mathrm{lin}}^\text{T} q^* \right).
\end{eqnarray}
Please note, the spatial and temporal integrals are not shown for the sake of simplicity.

The desired adjoint equation $E^* = E_{\mathrm{lin}}^\text{T}$ results from demanding
\begin{equation}
  g - E_{\mathrm{lin}}^\text{T} q^* = 0, \label{eq_adjoint_system}
\end{equation}
with $q^* = [\varrho^*,u^*_j,p^*]$ as adjoint state variable.

For a detailed derivation of the adjoint \textsc{Euler} equations see \cite{Lemke2015}.
They are given by
\begin{align}
  \partial_t q^* = \tilde A\, \left[ -({B^i})^\text{T} \partial_{x_i}q^* - \partial_{x_i} ({C^{i}})^\text{T} q^* + \tilde C^{i} \partial_{x_i} c - g \right] 
	\label{eq_adjoint_equation}
\end{align}
with $\tilde A = \left({A^\text{T}}\right)^{-1}$ and $\tilde C^{i}$ as resorting 
\begin{equation}
  q_\alpha^*\delta C^{i}_{\alpha\beta}\partial_{x_{i}} c_\beta = q_\alpha^* \delta q_\kappa \dfrac{\partial C_{\alpha\beta}^{i}}{\partial q_\kappa} \partial_{x_{i}} c_\beta
\end{equation}
abbreviated as $\delta q_\kappa \tilde C_{\kappa\beta}^{i} \partial_{x_{i}} c_\beta$.
The matrices $A$, $B^i$ and $C^i$ are given in the appendix.

Finally, the change of the objective function is given by
\begin{equation}
  \delta J = {q^*}^\text{T} \delta s.
\end{equation}
Thus, the solution of the adjoint equation can be interpreted as gradient of $J$ with respect to the source terms $s$
\begin{equation}
  \nabla_s J = {q^*}.
\end{equation}
Initial and boundary conditions of the adjoint Euler equations as well as the derivation of the adjoint compressible Navier-Stokes equations are discussed in \cite{Lemke2015}.

\subsection{Iterative Process}
%
\begin{figure*}[t!] 
  \centering
  \includegraphics[width=.8\textwidth]{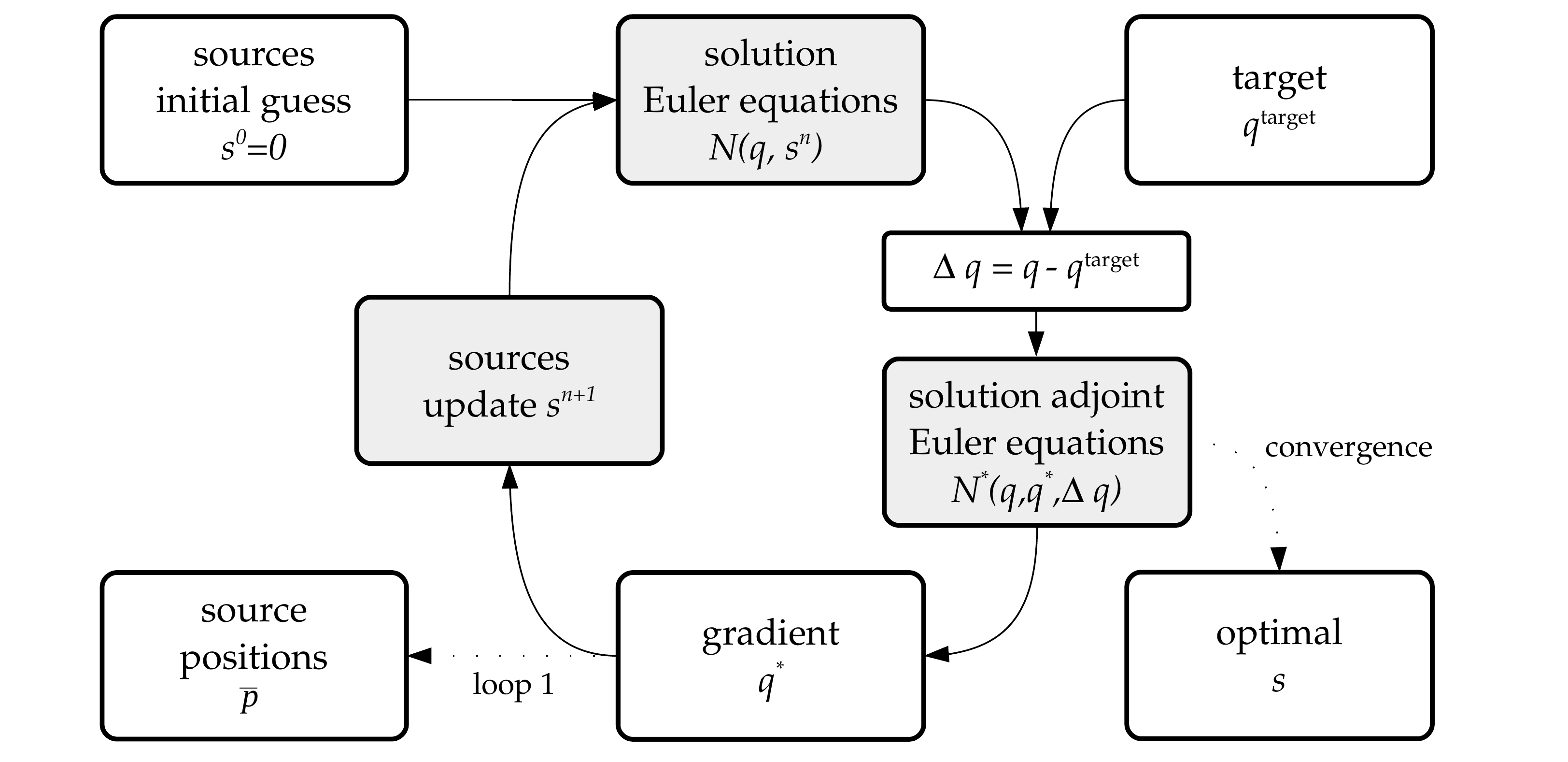}
  \caption{Iterative procedure for the determination of an optimal $s$. Computationally intensive steps are marked in gray. 
  The first gradient provides information on (optimal) source positions, see Sec.~\ref{sec_source_localization} for a detailed discussion.}
  \label{fig:GeneralLoop}
\end{figure*}

The adjoint-based gradient is employed in an iterative manner.
First, the Euler equations \eqref{eq_euler_equations_vectorwise} are solved forward in time, usually with $s^0 = 0$.
Subsequently, the adjoint equations \eqref{eq_adjoint_equation} are calculated backward in time, deploying the direct solution and $g$. 
Based on the adjoint solution, the gradient $\nabla_s J$ is determined and used to update the source distribution $s^n$ by means of a steepest gradient approach:
\begin{equation}
  s^{n+1} = s^n + \alpha \nabla_s J,	\label{eqn_steep_des}
\end{equation}
with $\alpha$ denoting an appropriate step size and $n$ the iteration number.
The gradient is calculated for the whole computing region and the entire simulation time.
For the determination of sound sources with a known position, the gradient is evaluated only there.
The procedure is repeated, using the current $s^n$, until a suitable convergence criterion is reached.
Typically, for acoustic problems, convergence is reached within or less 20 loops.

The identification of global optima is not ensured as the proposed technique optimizes to local extrema only.
The computational costs of the approach are independent of the number of sources and their arrangement.
However, they depend on the size and resolution of the computational domain in space and time, defined by the considered frequency range.
The computational problem is fully parallelizable.

\subsection{Source Localization}
In particular, when $s^0 = 0$ holds, the first adjoint solution contains information on the position of the sources.
By the pointwise summation of the absolute adjoint sensitivities $p^*$ in the spatial domain over all computed time steps
\begin{equation}
  \bar p = \sum \limits_{t_{n=0}}^{t_{n=\mathrm{end}}} | p^* |, \label{eq_sum_adjoint_sens}
\end{equation}
the positions featuring maximum impact on the objective function can be identified by means of maxima of $\bar p$.
These correspond to the most likely (monopole) source locations.
Thus, the adjoint solution allows the localization of sound sources, see Sec.~\ref{sec_source_localization}.
A subsequent iterative adaptation of the sources can be interpreted as adjoint-based monopole synthesis.

\section{Adjoint CAA framework \label{sec_caa_framework}}
\begin{figure}
    \centering
    \includegraphics[width = .49\textwidth]{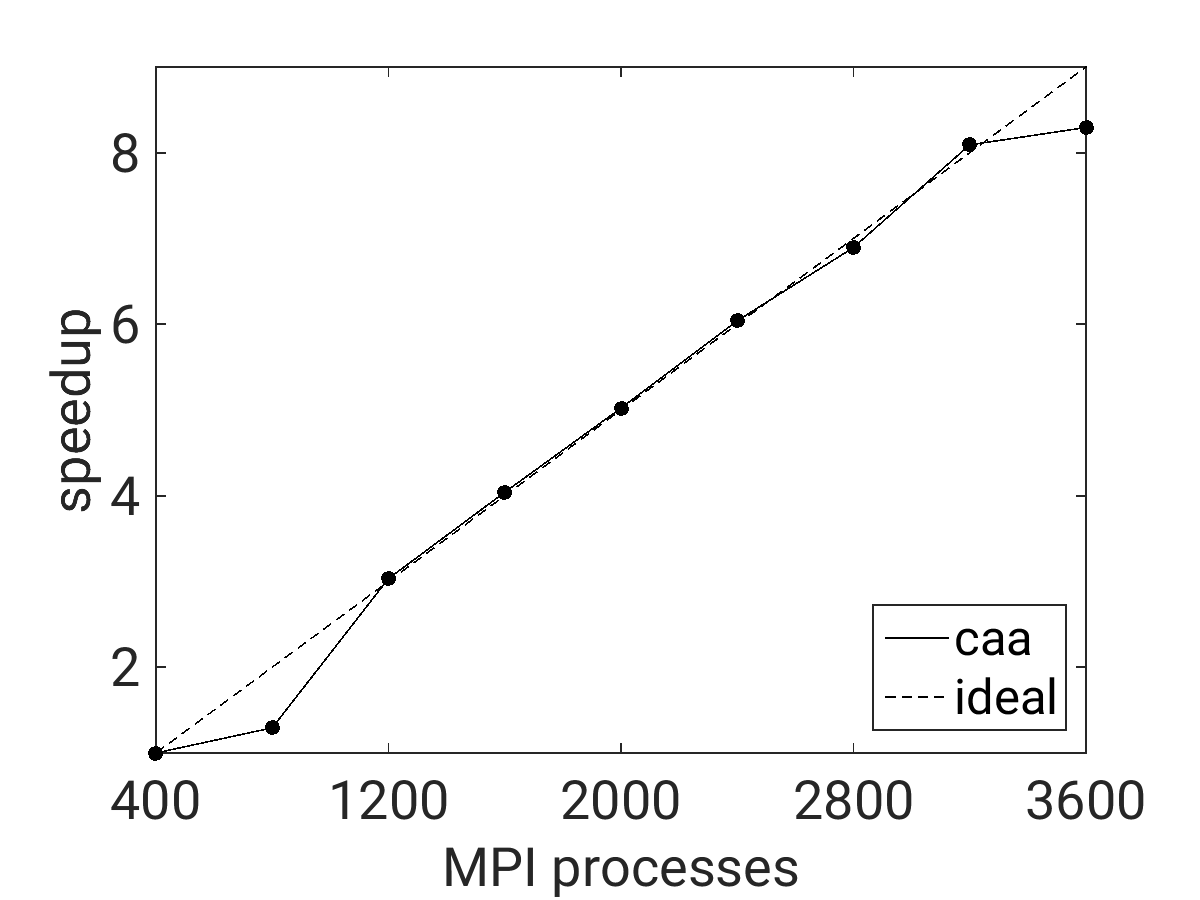}
    \includegraphics[width = .49\textwidth]{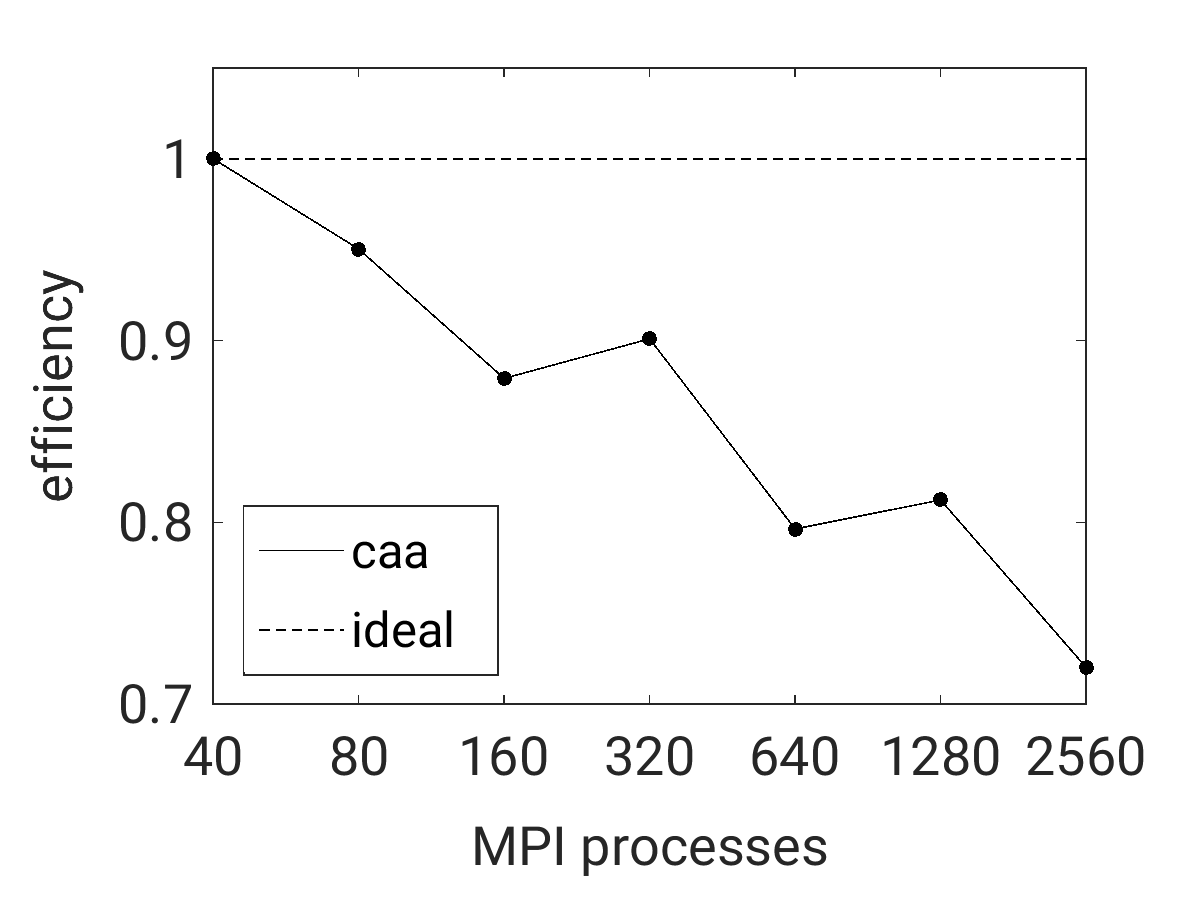}
    \caption{(Left) Strong scaling behaviour. 
    The overall number of grid points is kept constant while increasing the number of MPI processes. 
    Nearly linear scaling is found.
    (Right) Weak scaling behaviour. 
    The number of grid points on each process is kept constant while increasing the number of MPI processes.
    An admissible reduction of the parallelization efficiency is found.}
    \label{fig_scaling}
\end{figure}

The set of governing equations \eqref{eq_euler_equations_vectorwise} is implemented by means of a new MPI-parallelized Fortran program.
The discretization is realized by a finite difference time domain approach (FDTD).
For the spatial derivatives, a compact scheme of 6th order is employed \cite{Lele1992}.
The corresponding linear system of equations is solved by BLAS routines using an LU-decomposition.
For the time-wise integration, the standard explicit \textsc{Runge-Kutta}-scheme of fourth-order is used.
To ensure stability, a compact filter is employed \cite{GaitondeVisbal2000}.
Boundaries are treated by characteristic boundary conditions \cite{PoinsotLele1992}.
The MPI implementation is realized by collective communication via all2all\_v.
The parallelization strategy is found to be efficient for the governing equations \eqref{eq_euler_equations_vectorwise}, see Fig.~\ref{fig_scaling}, and comparable to other implementations using collective communication, e.g.~\cite{Pekurovsky2012}.

Thus, the code is prepared to handle large scale problems, e.g.,~open-air festival sites in the context of sound reinforcement applications or source localization for vehicle aeroacoustics in wind tunnels.
However, the examples presented in the following are computed using a single workstation or a few cluster nodes.

The adjoint equations are solved using the same discretization.
A detailed discussion on the adjoint initial- and characteristic boundary conditions can be found in \cite{Lemke2015}.

\section{Application I: Sound Reinforcement \label{sec_sound_reinforcement}}
%
This section presents a test case regarding the optimization of sound reinforcement setups.
The overall goal is to identify optimal drives (amplitude and phase) for given loudspeakers in order to synthesise a desired sound field.
The loudspeakers are approximated by means of monopole sources, which is feasible for low frequencies.

The spatial domain under consideration is $1.6 \times 1.6 \times 1.6$ m$^3$.
The domain is resolved by $197 \times 197 \times 99$ equidistantly distributed points.
The time step, and by this, the sampling rate, is given by 48 kHz, corresponding to a CFL-condition smaller than 1.
The computational time span considered is 31.25 ms.
The reference values for density and pressure correspond to a speed of sound of 343 m/s.
All boundaries are treated as non-reflecting.
In addition, a sponge layer is applied at all boundaries.

\begin{figure}
    \centering
    \includegraphics[width = .49\textwidth]{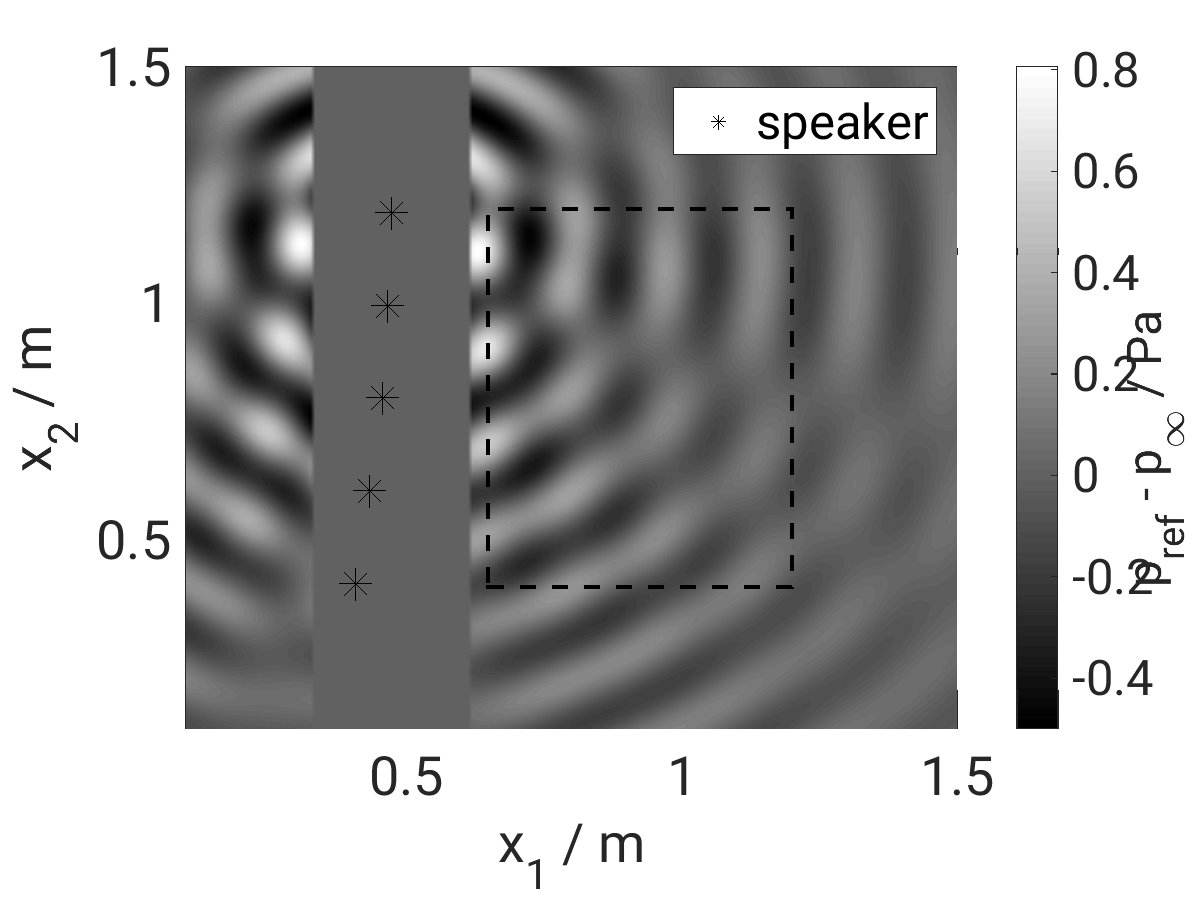}
    \includegraphics[width = .49\textwidth]{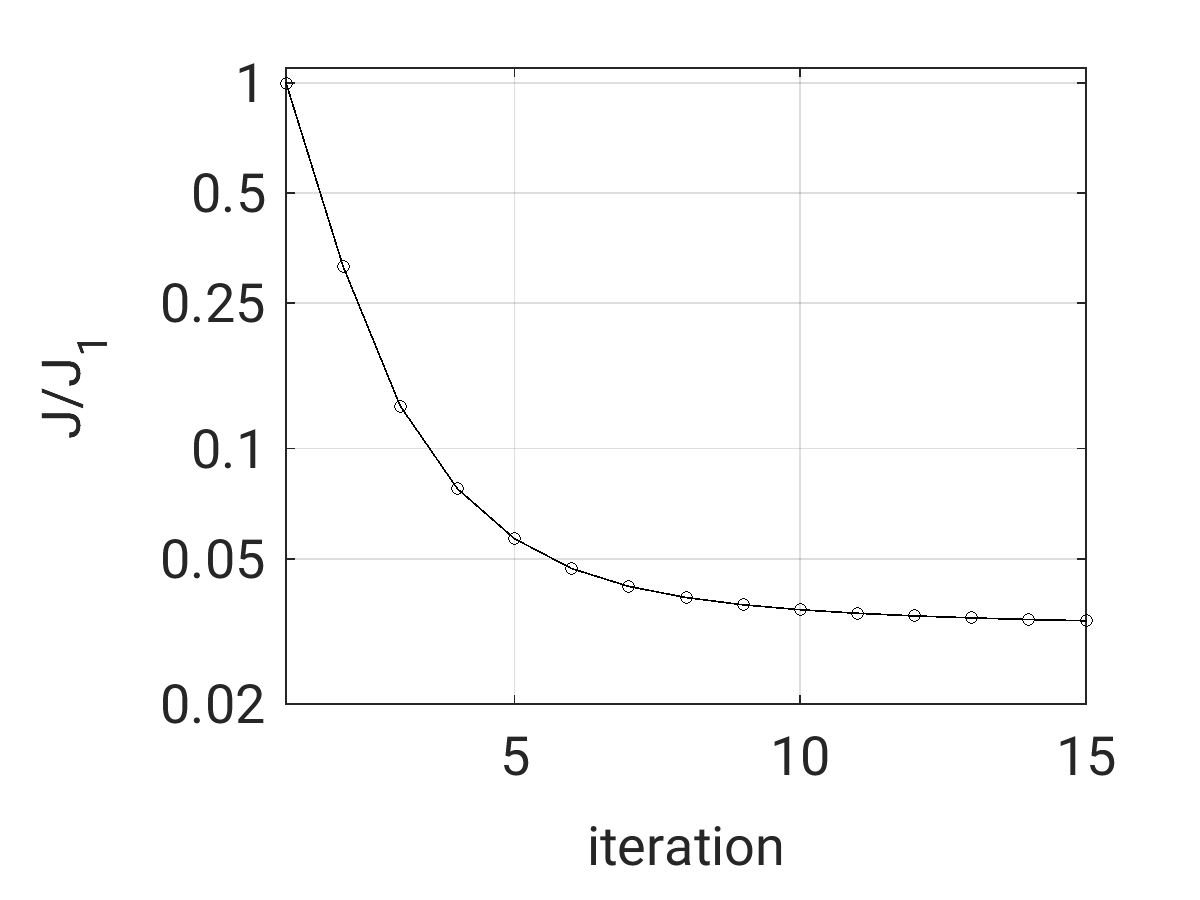}
    \caption{(Left) Sound reinforcement setup including a selected time step of the CDPS-based reference sound field shown at the center $x_1$-$x_2$ plane of the computational domain.
    The five monopole speakers in a curved arrangement are denoted by (*).
    Different driving functions (in amplitude and phase) for the speaker result in a steered sound field.
    The area/volume marked by the dashed line corresponds to the spatial weight $\sigma$ in the objective function.
    Please note, the employed CDPS technique for computing the reference sound field does not provide reliable solutions near the source positions; therefore, $p'_{\mathrm{ref}}$ is discontinuous for $x_1 = [0.32,0.62]$ m.
    (Right) Progress of the objective function with a logarithmic y-axis.
    Convergence is reached.
    The objective is reduced by nearly two orders of magnitude with respect to the initial guess $s=0$.}
    \label{fig_reinf1_setup_objective}
\end{figure}

For the test case reference signals for five sources, located in a curved arrangement in the center $x_1$-$x_2$ plane, are predefined.
The signals are characterised by different amplitudes and phase delays resulting in a steered sound field, see Fig.~\ref{fig_reinf1_setup_objective} (left).
In order to investigate the frequency band 1-3 kHz, a corresponding logarithmic sine-sweep is specified as the reference signal.
Using this setup, a reference sound field is computed by a Complex Directivity Point Source (CDPS) algorithm \cite{Feistel2014}.
The resulting reference sound field serves as the target for the adjoint-based framework, with the aim to identify the reference signals (amplitudes and phases) based on the reference target sound field only.

\begin{figure}
    \centering
    \includegraphics[width = .49\textwidth]{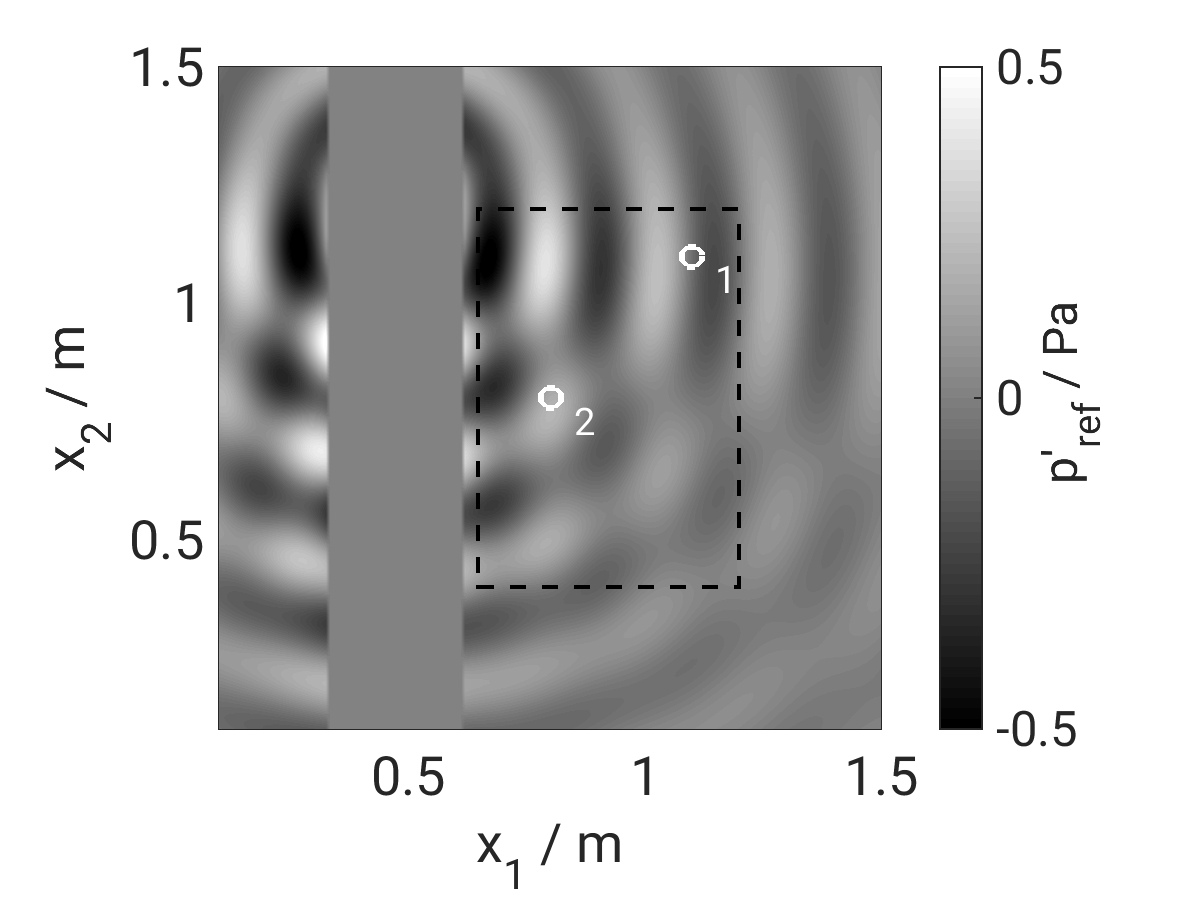}
    \includegraphics[width = .49\textwidth]{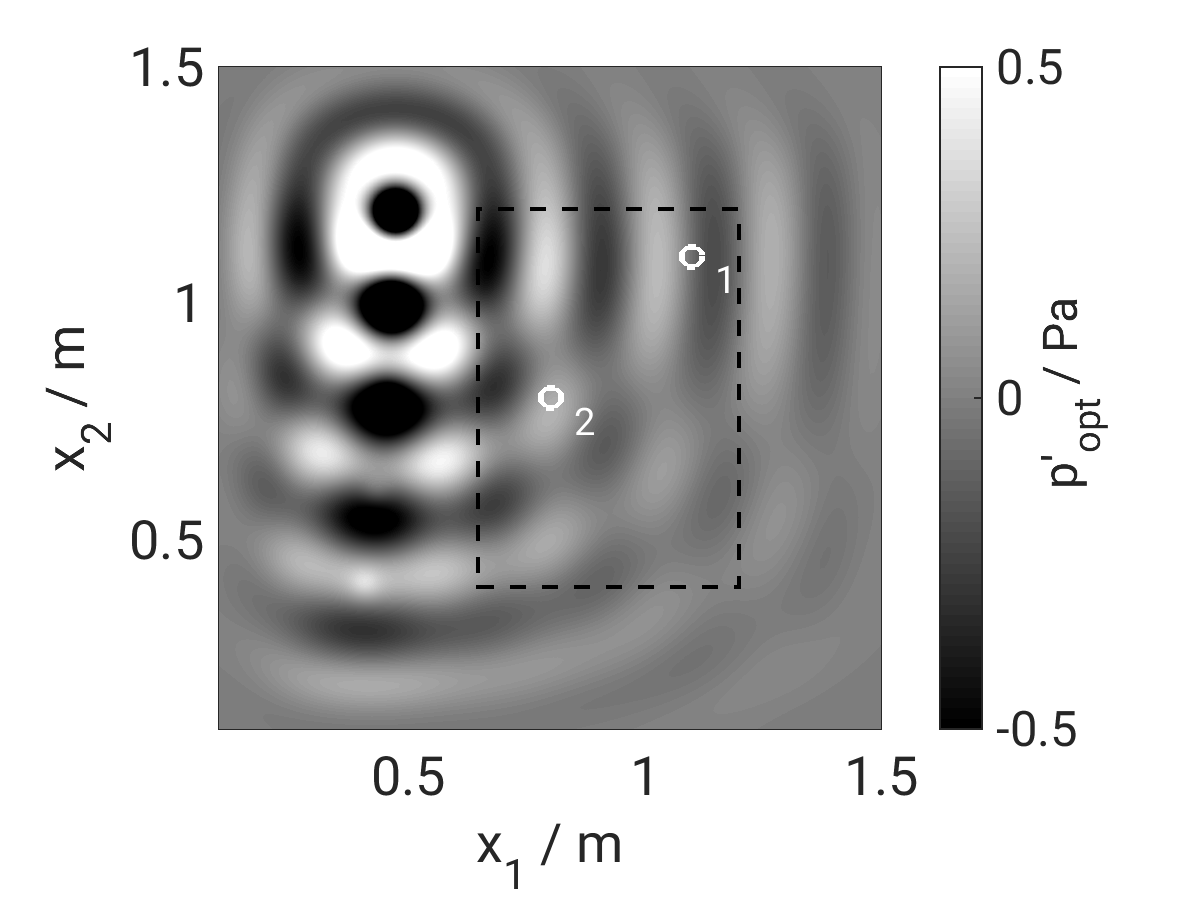}
    \caption{Reference target (left) and resulting optimized (right) sound field at $t = 15.63$ ms for the center $x_1$-$x_2$ plane.
    The general features of the reference field are (re-) captured.
    The influence of the employed sponge layer in the adjoint-based sound field is visible.
    The dashed line encodes the spatial weight $\sigma$ within the objective function.
    The marked positions correspond to synthetic microphone positions $x_{1,2} = [1.1,1.1]$ and $x_{1,2} = [0.8,0.8]$ which are used for spectral analysis, see text for details.}
    \label{fig_reinf1_field}
\end{figure}

After 15 iterative loops of the adjoint framework, the objective function is reduced to nearly 3\% with respect to the initial solution with $s=0$, see Fig.~\ref{fig_reinf1_setup_objective} (right).
The general features of the target reference sound field are captured, see Fig.~\ref{fig_reinf1_field}.
A detailed spectral analysis of the occurring deviations at two selected microphone positions, presented in Fig.~\ref{fig_reinf1_field}, show amplitude deviations less than 1 dB within the confidence interval from 1.3 to 2.7 kHz.
The normalized phase derivations, with respect to $2\pi$, are in the limits of -0.07 to 0.07.
\begin{figure}
    \centering
    \includegraphics[width = .49\textwidth]{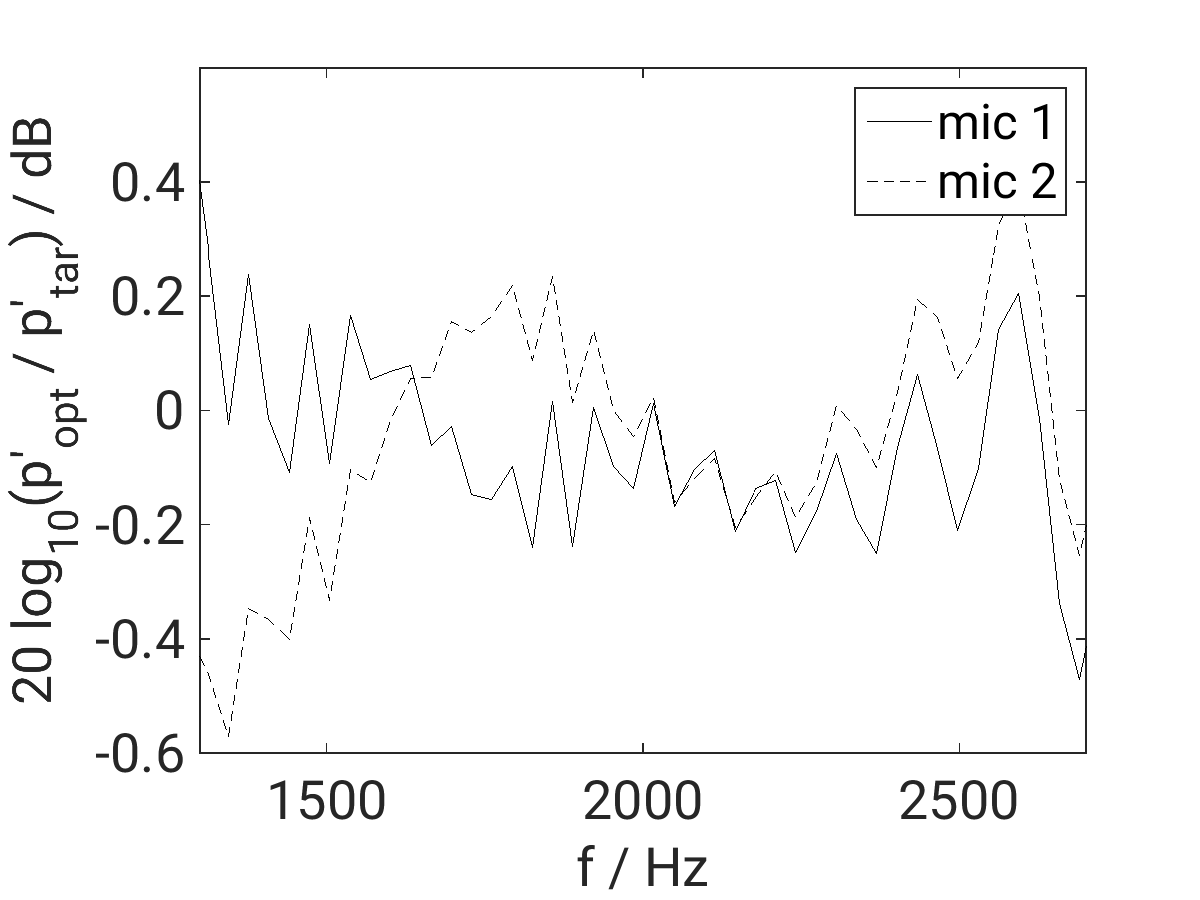}
    \includegraphics[width = .49\textwidth]{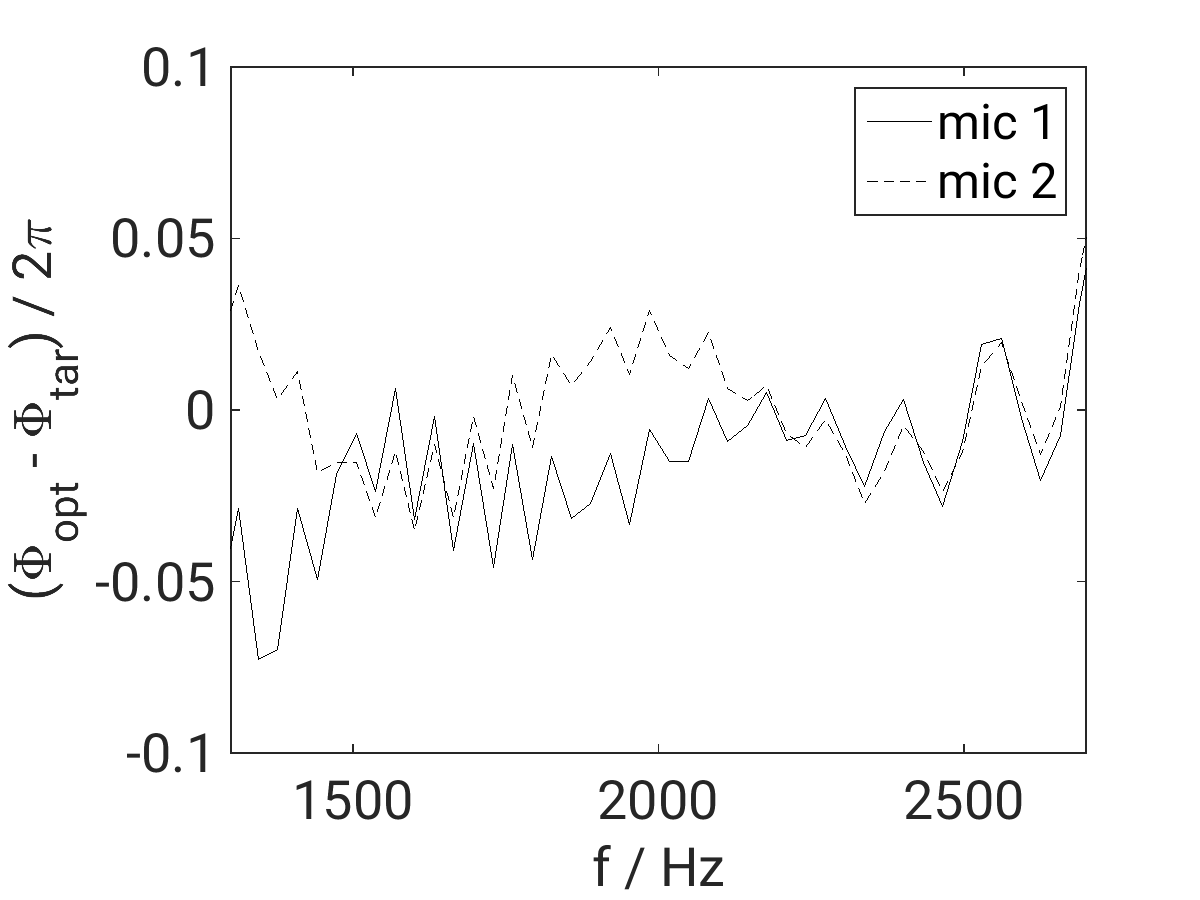}
    \caption{(Left) Normalized amplitude difference between resulting optimized and reference target sound field at selected microphone positions, see Fig.~\ref{fig_reinf1_field}.
    (Right) Normalized phase difference between resulting optimized and reference target sound field at the selected microphone positions.
    \label{fig_reinf2_delta}}
\end{figure}

A discussion on how to derive optimal electronic drives from the adjoint-based signals $s$ is given in \cite{SteinStraubeSesterhennWeinzierlLemke2019}. 
Therein, the capability of the approach to consider complex base flows by means of wind and temperature stratification is shown.

\section{Application II: Source Localization \label{sec_source_localization}}
%
In this section, the localization of fixed and moving sound sources is shown.
Two generic setups serve as a proof of concept.
For the first setup with four stationary sound sources and the second setup with a moving source, it is shown that the adjoint-based approach is able to identify the sources and track their path in case of moving.

In both cases, the measurements are provided by a reference computation with predefined sound sources.
Synthetic microphone signals are extracted from this reference solution.
A spatially discrete planar array with 64 microphones is used.
The general setup is based on the array benchmark test case B7 provided by the Brandenburg university of technology, see \cite{Geyer2019}.
Modifications are discussed below.
An example in which experimental data are used is shown in \cite{Lemke2015}.

The spatial domain under consideration is $1.7 \times 1.7 \times 1.25$ m$^3$. 
The domain is resolved by $240 \times 240 \times 176$ equidistantly distributed points.
The time step, and by this, the sampling rate of the microphone measurements, is given by 53.33 kHz, corresponding to a CFL-condition smaller than 1.
In both cases, no base flow is considered.
The reference values for density and pressure correspond to a speed of sound of 343 m/s.
The spiral-like microphone array is located at $x_3 = 0$ m and centered in the corresponding plane.
The spatial distribution of the microphones is described in more detail in \cite{Geyer2019}.
All boundaries are treated as non-reflecting.
In addition, a sponge layer is applied at all boundaries.

\subsection{Four sources}
\begin{figure}
    \centering
    \includegraphics[width = .49\textwidth]{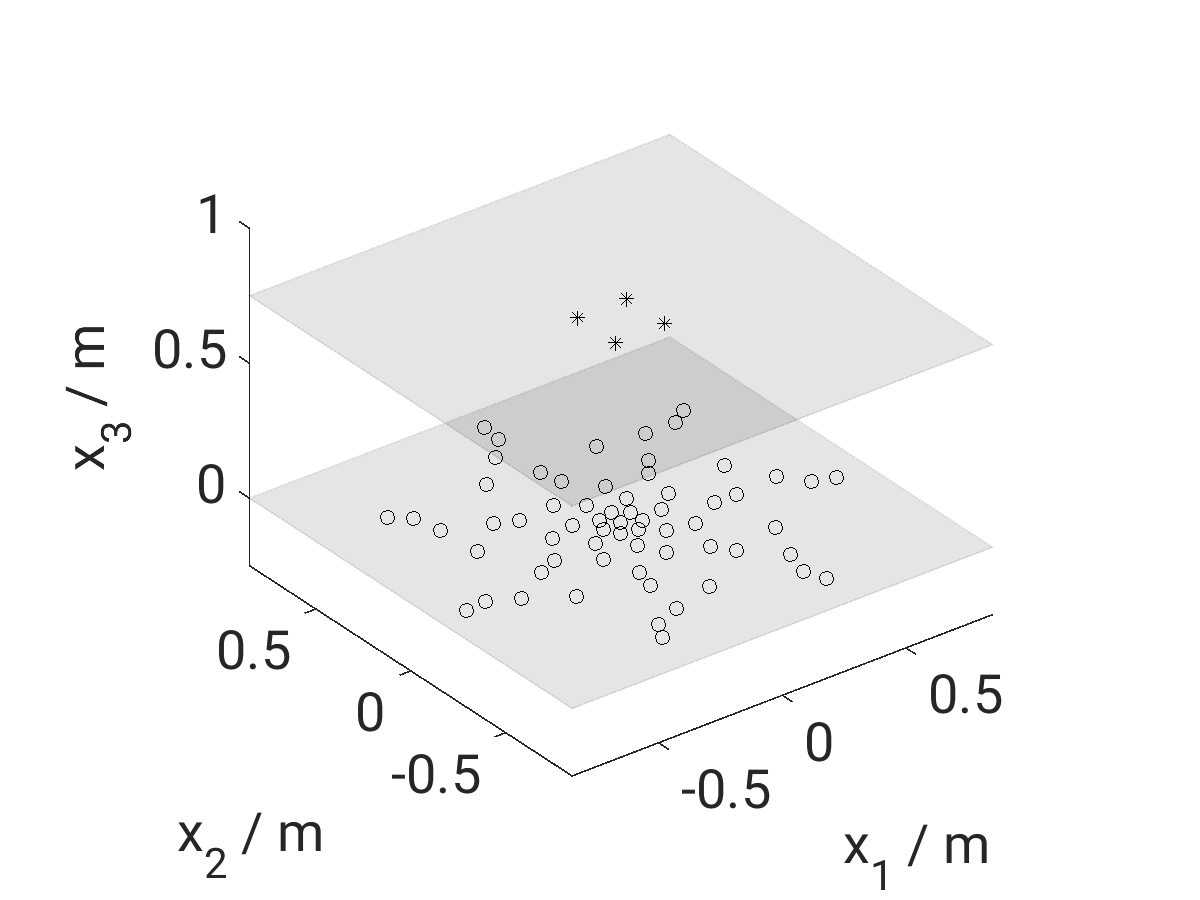}
    \includegraphics[width = .49\textwidth]{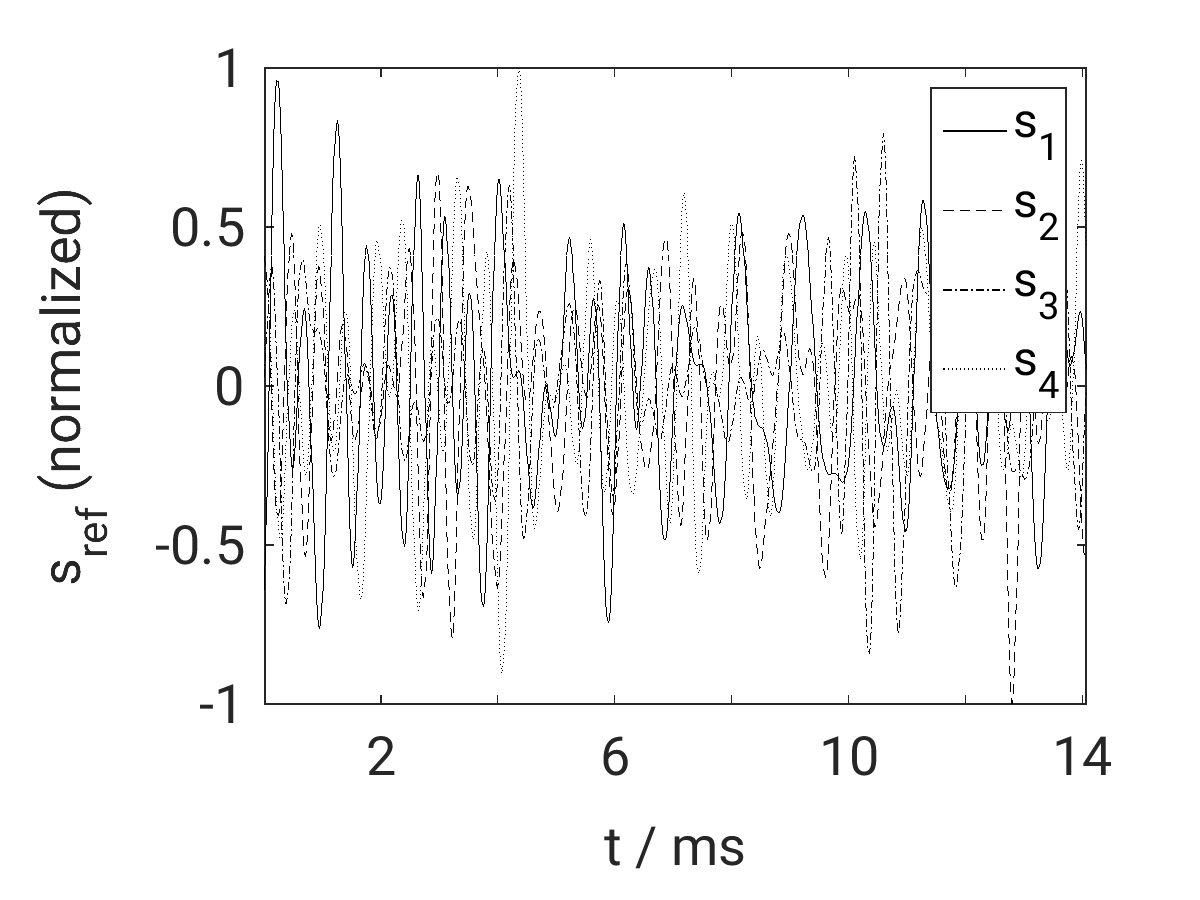}
    \caption{(Left) Acoustic setup for source localization of four sources (*) by 64 microphones (o) located in the planes $x_3=0.75$ m respectively $x_3=0$ m. (Right) Normalized signals $s_i$ of the four reference sources, shown for the whole computational time.}
    \label{fig_loc1_setup_s_signals}
\end{figure}

As in the array benchmark test case B7 four monopole sources are located in the $x_1$-$x_2$-plane at $x_3 = 0.75$ m, see Fig.~\ref{fig_loc1_setup_s_signals} (left).
For the reference computation, the original benchmark source signals are replaced by incoherent random signals, frequency-band limited between 750 and 2500 Hz, see Fig.~\ref{fig_loc1_setup_s_signals} (right).
The computational time span is 14.06 ms. 

\begin{figure}
    \centering
    \includegraphics[width = .49\textwidth]{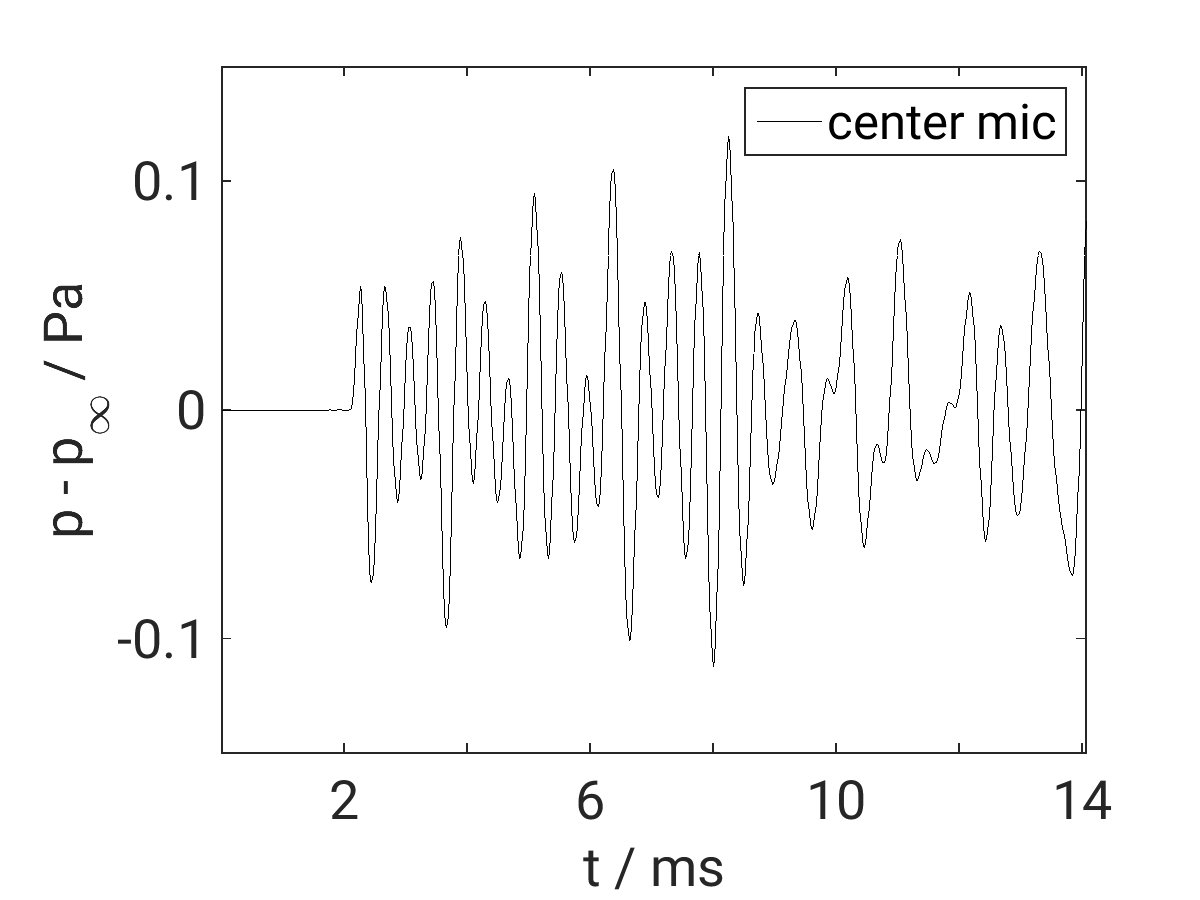}
    \includegraphics[width = .49\textwidth]{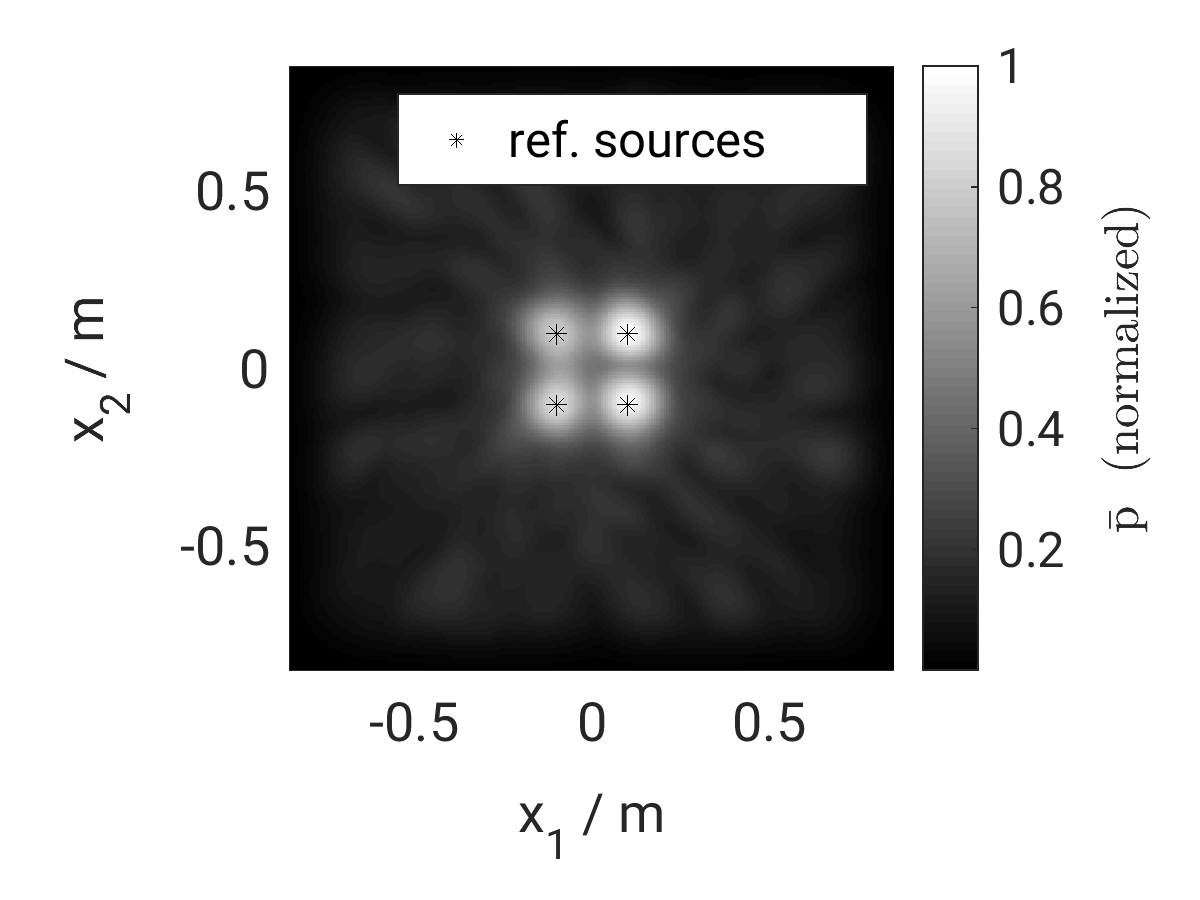}
    \caption{(Left) Captured pressure signal at the center microphone in the array.
    The initial silence results from the distance between the sources and the array.
    (Right) Resulting pointwise summation of the absolute adjoint sensitivities $p^*$ \protect\eqref{eq_sum_adjoint_sens}.
    The reference source positions ($*$) are recovered.}
    \label{fig_loc1_center_mic_p_hat_plane}
\end{figure}

Using a corresponding reference forcing $s = \sum_{i} s_i$ a simulation of the \textsc{Euler} equations \eqref{eq_euler_equations_vectorwise} is carried out.
From the results, discrete microphone signals are extracted, see Fig.~\ref{fig_loc1_center_mic_p_hat_plane} (left), which are the result of the superposition of all sources and the associated signals.

The 64 signals are encoded in the objective function $J$ \eqref{eqn_objective} using the spatial weight $\sigma$.
To avoid an unstable discrete forcing of the adjoint equations, $\sigma$ is chosen as \textsc{Gauss}-distribution with a half-width of $2\Delta x$ for each microphone position.
After determining the solution of the direct equations with an initial guess for $s = 0$, here, constant environmental conditions for all time steps, the adjoint equations are solved backwards in time.
From the resulting gradient, the source positions can be derived, as discussed before.
That way, the reference source positions are identified, see Fig.~\ref{fig_loc1_center_mic_p_hat_plane} (right).

Please note, the analysis is based on the first adjoint-based gradient only.
The required computational time for the analysis is less than 15 min on a 16 core workstation.
Iterative optimization of $s$ might improve the results.

\subsection{Single moving source}
\begin{figure}
    \centering
    \includegraphics[width = .49\textwidth]{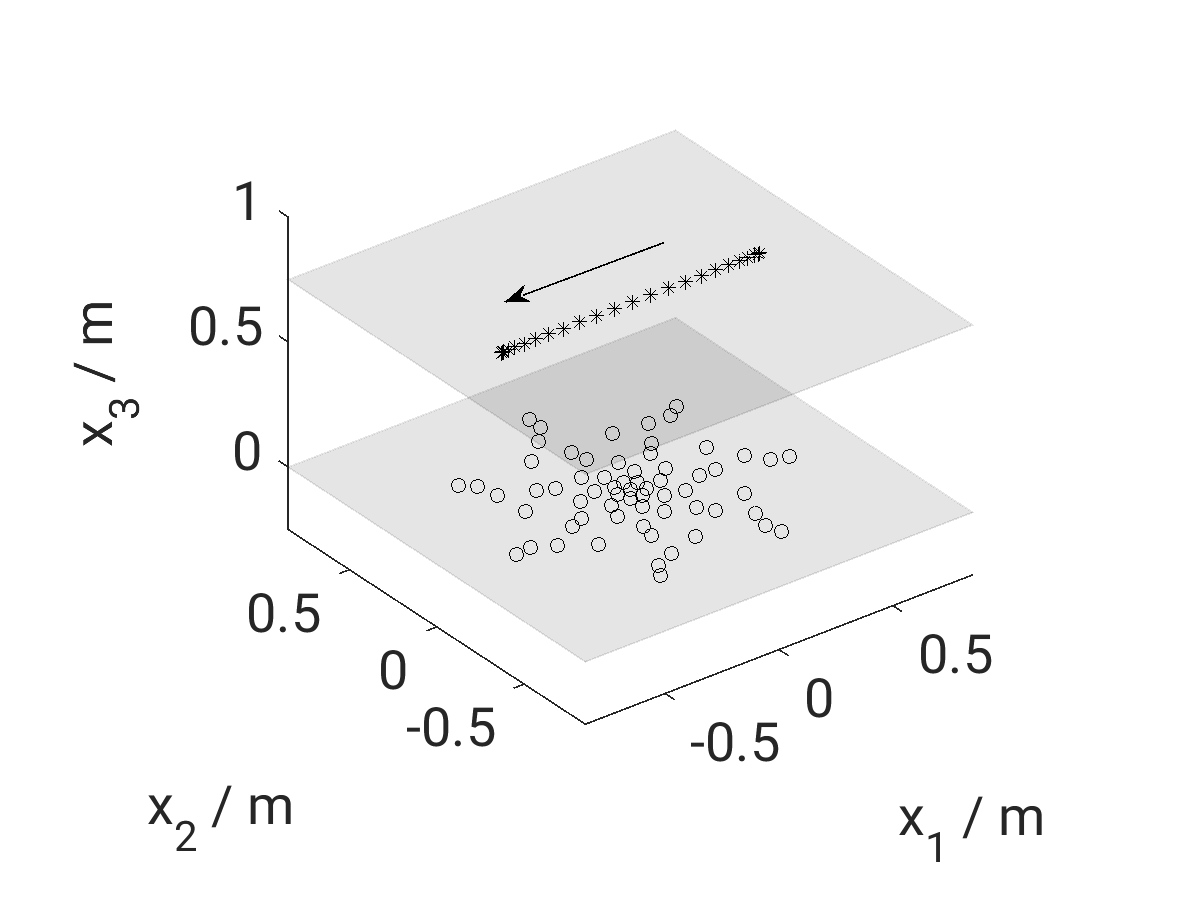}
    \includegraphics[width = .49\textwidth]{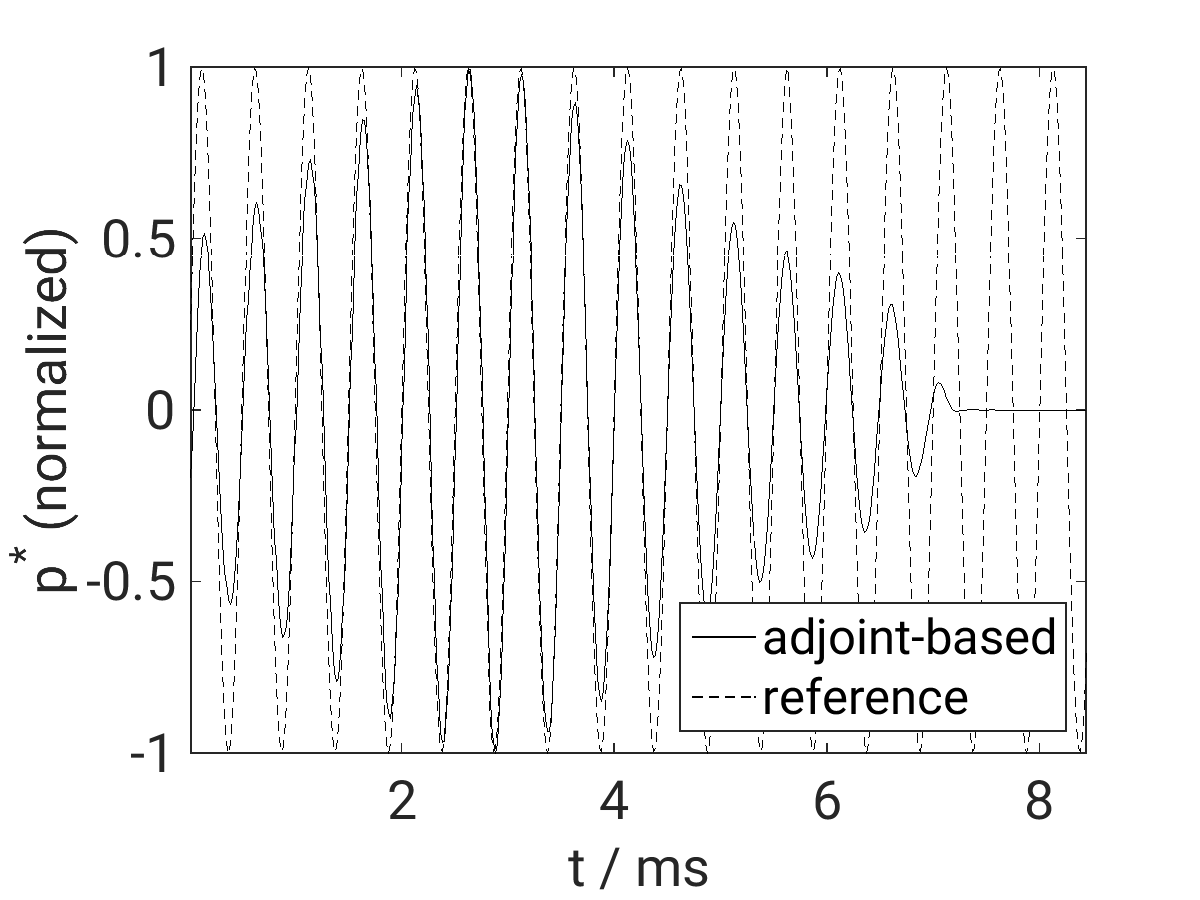}
    \caption{(Left) Acoustic setup for source localization of a single moving source (*) by means of 64 microphones (o) located in the planes $x_3=0.75$ m, respectively $x_3=0$ m.
    The movement of the source is visualized by it waypoints, chosen with a constant time interval.
    (Right) Normalized adjoint-based sensitivity $p^*$ at the reference source positions over time in comparison to the reference forcing.
    See text for a detailed discussion.}
    \label{fig_loc2_setup_s_signal}
\end{figure}

Again, the aforementioned test case B7 from \cite{Geyer2019} serves as a base for the following test setup.
The planar microphone array is located in the same plane ($x_3 = 0$) but scaled by a factor of 0.8, resulting in smaller distances between the microphones.
The incoherent sources are replaced by a single source with a harmonic 2 kHz reference signal.
The source is moving in the $x_1$-$x_2$-plane, see Fig.~\ref{fig_loc2_setup_s_signal} (left).
The movement is described by an acceleration and deceleration, taking place along the $x_1$ axis.
It starts at the beginning of the computational time and ends with the simulation after 8.44 ms. 
The highest speed of the movement is reached midway.

Again a reference solution provides synthetic microphone signals, which are encoded in the objective function.
Using constant environmental conditions as solution of the direct equations ($s^0=0$), the adjoint equations are solved.
Evaluation of the adjoint sensitivity $p^*$ over time at the reference source position provides information of the reference signal, see Fig.~\ref{fig_loc2_setup_s_signal} (right).
The phase of the reference signal is determined with very good agreement.
The amplitude shows deviations at the beginning and end of the simulation.
The influence of the directional characteristic of the used microphone array is presumed.

\begin{figure}
    \centering
    \includegraphics[width = .49\textwidth]{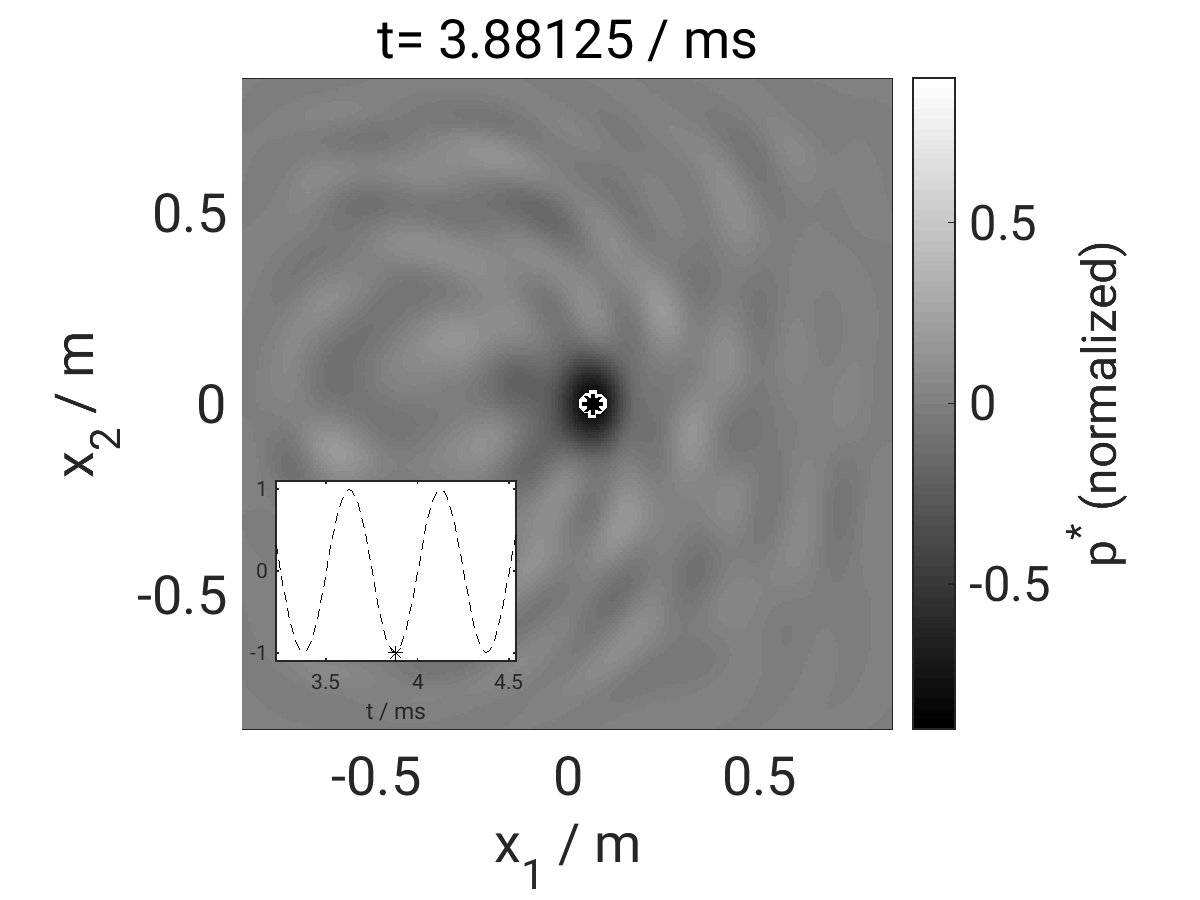}
    \includegraphics[width = .49\textwidth]{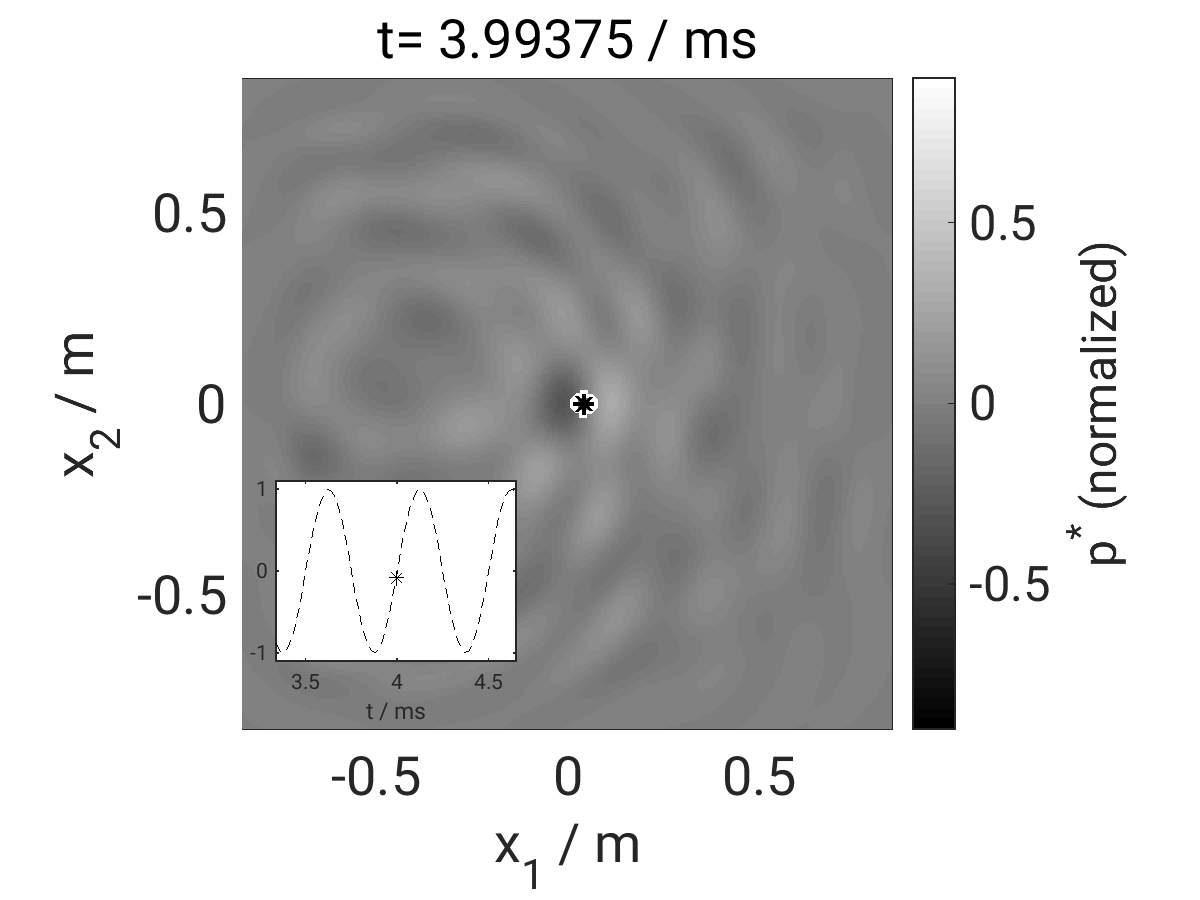} \\
    \includegraphics[width = .49\textwidth]{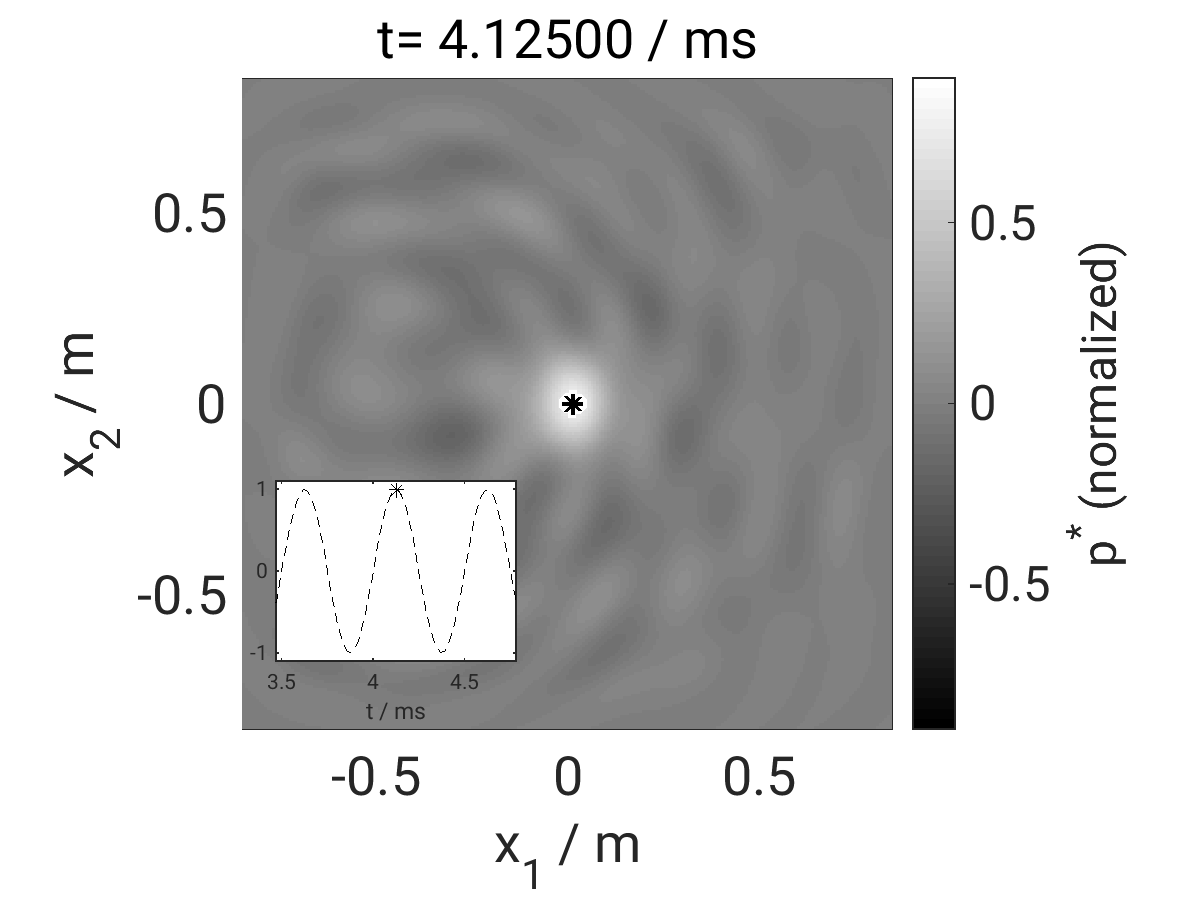}
    \includegraphics[width = .49\textwidth]{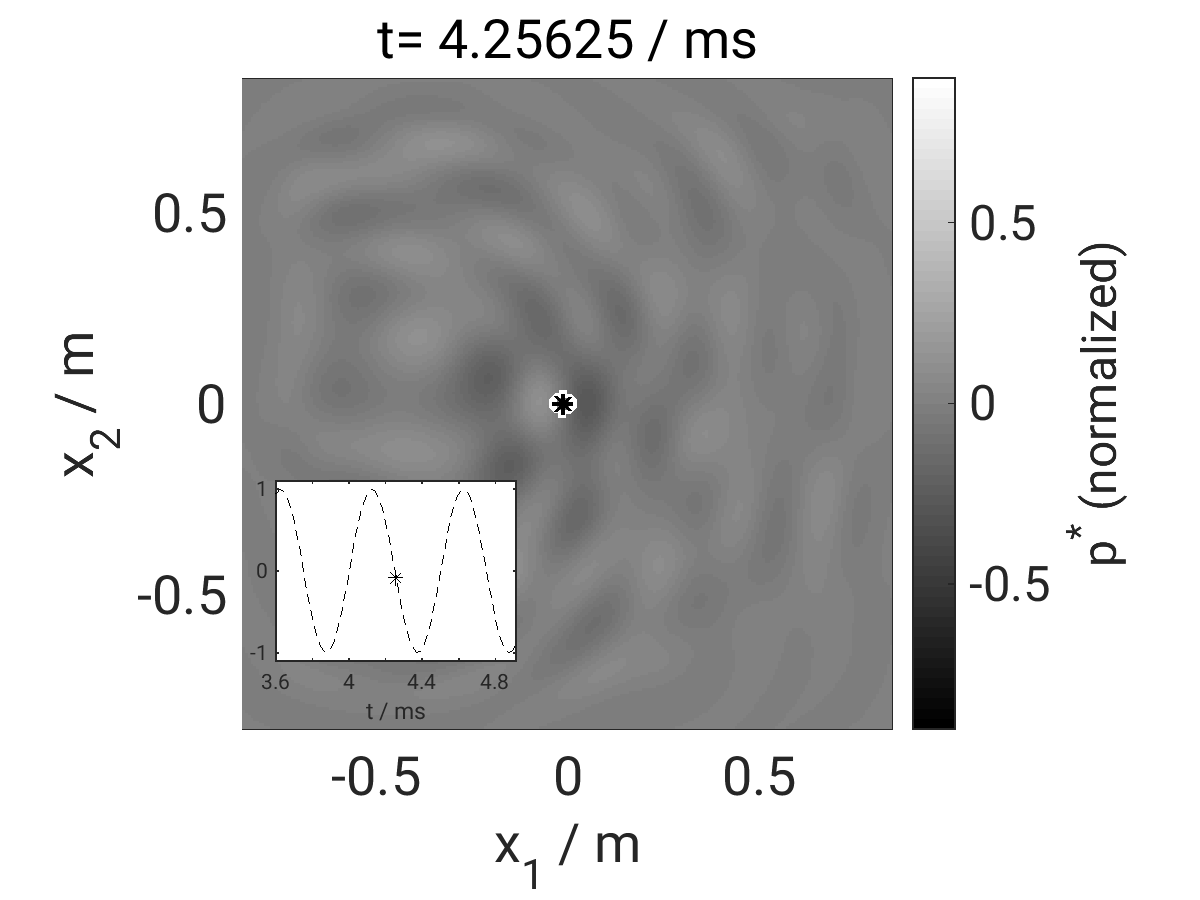}
    \caption{Normalized adjoint-based sensitivity $p^*$ at the plane $x_3 = 0.75$ for different time steps.
    The reference source location is marked by (*) in a white circle.
    In the inset, the normalized reference signal is shown.}
    \label{fig_loc2_ps_in_plane}
\end{figure}

Besides, the identification of the source signal also its position might be tracked.
In Fig.~\ref{fig_loc2_ps_in_plane} the adjoint-based sensitivity $p^*$ is shown for the plane $x_3 = 0.75$ m for different time steps.
Occurring maxima give rise to the actual sound source position, besides its signal.

Again, the analysis is based on the first adjoint-based gradient only.
The required computational time for the analysis is less than 10 min on 8 cluster nodes with 8 cores each.

\section{Summary}
%
An adjoint-based framework for the identification of sound sources is presented.
It is shown that the approach is able to determine (optimal) source signals and to track moving sources.

By design, the time-domain approach allows the consideration of base flows, such as velocity profiles and temperature stratification, and complex geometries, which will be the focus of the upcoming work.
The first results that take into account a complex base flow in the context of sound reinforcement are shown in \cite{SteinStraubeSesterhennWeinzierlLemke2019}.

\section*{Acknowledgments}
The authors gratefully acknowledge financial support by the Deutsche Forschungsgemeinschaft (DFG) within the project LE 3888/2-1.

We thank Florian Straube (Audio Communication Group, TU Berlin) for defining the target sound field for the sound reinforcement test case.

%
%
\bibliographystyle{abbrv}
\bibliography{bibdata}

\appendix
\section{\label{sec:appendix} Appendix}
\subsection{Adjoint equations}
As stated above, linearization of the governing \textsc{Euler} equations with respect to all state variables by $q = q_0 + \delta q$ results in
\begin{align}
      \partial_t 	A 		\delta q
   + \partial_{x_i}	B^{i} 		\delta q
    + C^{i} 		\partial_{x_i} 	\delta q
    + \delta C^{i} 	\partial_{x_i} 	c
  = 
    \delta s. \label{eq_ns_linear_quasi_linear_form_app}
\end{align}
Again, the summation convention applies.
The corresponding linearization matrices are

\begin{flalign*}
  & A = \left[\begin{matrix}1 & 0 & 0 & 0 & 0\\u_{1} & \rho & 0 & 0 & 0\\u_{2} & 0 & \rho & 0 & 0\\u_{3} & 0 & 0 & \rho & 0\\0 & 0 & 0 & 0 & \frac{1}{\gamma -1}\end{matrix}\right]\,, &
  & B^{1} = \left[\begin{matrix}u_{1} & \rho & 0 & 0 & 0\\u_{1}^{2} & 2 \rho u_{1} & 0 & 0 & 1\\u_{1} u_{2} & \rho u_{2} & \rho u_{1} & 0 & 0\\u_{1} u_{3} & \rho u_{3} & 0 & \rho u_{1} & 0\\0 & \frac{\gamma p}{\gamma -1} & 0 & 0 & \frac{\gamma u_{1}}{\gamma -1}\end{matrix}\right]\,, &
\end{flalign*}
\begin{flalign*}
   & B^{2} = \left[\begin{matrix}u_{2} & 0 & \rho & 0 & 0\\u_{1} u_{2} & \rho u_{2} & \rho u_{1} & 0 & 0\\u_{2}^{2} & 0 & 2 \rho u_{2} & 0 & 1\\u_{2} u_{3} & 0 & \rho u_{3} & \rho u_{2} & 0\\0 & 0 & \frac{\gamma p}{\gamma -1} & 0 & \frac{\gamma u_{2}}{\gamma -1}\end{matrix}\right]\,, &
  & B^{3} = \left[\begin{matrix}u_{3} & 0 & 0 & \rho & 0\\u_{1} u_{3} & \rho u_{3} & 0 & \rho u_{1} & 0\\u_{2} u_{3} & 0 & \rho u_{3} & \rho u_{2} & 0\\u_{3}^{2} & 0 & 0 & 2 \rho u_{3} & 1\\0 & 0 & 0 & \frac{\gamma p}{\gamma -1} & \frac{\gamma u_{3}}{\gamma -1}\end{matrix}\right]\,, &
\end{flalign*}
\begin{flalign*}
  & C^{i} = \left[\begin{matrix}0 & 0 & 0 & 0 & 0\\0 & 0 & 0 & 0 & 0\\0 & 0 & 0 & 0 & 0\\0 & 0 & 0 & 0 & 0\\0 & 0 & 0 & 0 & - u_{i}\end{matrix}\right]\,,\quad
  \delta C^{i} = \left[\begin{matrix}0 & 0 & 0 & 0 & 0\\0 & 0 & 0 & 0 & 0\\0 & 0 & 0 & 0 & 0\\0 & 0 & 0 & 0 & 0\\0 & 0 & 0 & 0 & - \delta u_{i}\end{matrix}\right]\,. &
\end{flalign*}

The full adjoint \textsc{Navier-Stokes} equations, in particular, the friction terms, are derived and discussed in \cite{Lemke2015}.
The two-dimensional adjoint \textsc{Euler} equations can be found in \cite{LemkeStraubeSchultzSesterhennWeinzierl2017}.

\vfill

\noindent status: draft for review \\
\noindent last modified: \today{} by (ML)

\end{document}